\documentclass[onecolumn]{cinc}
\usepackage{graphicx}

\usepackage{comment}

\usepackage{xcolor}

\usepackage{multirow}
\usepackage[normalem]{ulem} 
\newcommand{\stkout}[1]{\ifmmode\text{\sout{\ensuremath{#1}}}\else\sout{#1}\fi}

\usepackage{longtable}

\usepackage{hyperref}

\begin{document}
\bibliographystyle{cinc}

\title{Explainable and externally validated machine learning for neurocognitive
diagnosis via electrocardiograms}

\author{
Juan Miguel Lopez Alcaraz\textsuperscript{1,*},  
Ebenezer Oloyede\textsuperscript{2,3},  
David Taylor\textsuperscript{4},  
Wilhelm Haverkamp\textsuperscript{5},  
Nils Strodthoff\textsuperscript{1,$\dagger$}
}

\maketitle

\begin{center}
\textsuperscript{1}AI4Health Division, Carl von Ossietzky Universität Oldenburg, Oldenburg, Germany. \\
\textsuperscript{2}Pharmacy Department, Maudsley Hospital, London, UK. \\
\textsuperscript{3}Department of Psychiatry, University of Oxford, UK. \\
\textsuperscript{4}Institute of Pharmaceutical Sciences, King's College London, UK. \\
\textsuperscript{5}Department of Cardiology, German Heart Center of the Charité-University Medicine, Charité Campus Mitte, Berlin, Germany. \\
\textsuperscript{*}First author\\
\textsuperscript{$\dagger$}Senior and corresponding author: nils.strodthoff@uol.de
\end{center}

\begin{abstract}

\textbf{Background: } Electrocardiogram (ECG) analysis has emerged as a promising tool for detecting physiological changes linked to non-cardiac disorders. Given the close connection between cardiovascular and neurocognitive health, ECG abnormalities may be present in individuals with co-occurring neurocognitive conditions. This highlights the potential of ECG as a biomarker to improve detection, therapy monitoring, and risk stratification in patients with neurocognitive disorders, an area that remains underexplored.

\textbf{Methods: } We aim to demonstrate the feasibility to predict neurocognitive disorders from ECG features across diverse patient populations. We utilized ECG features and demographic data to predict neurocognitive disorders defined by ICD-10 codes, focusing on dementia, delirium, and Parkinson’s disease. Internal and external validations were performed using the MIMIC-IV and ECG-View datasets. Predictive performance was assessed using AUROC scores, and Shapley values were used to interpret feature contributions.

\textbf{Results: } Significant predictive performance was observed for disorders within the neurcognitive disorders. Significantly, the disorders with the highest predictive performance is F03: Dementia, with an internal AUROC of 0.848 (95\% CI: 0.848-0.848) and an external AUROC of 0.865 (0.864-0.965), followed by G30: Alzheimer's, with an internal AUROC of 0.809 (95\% CI: 0.808-0.810) and an external AUROC of 0.863 (95\% CI: 0.863-0.864). Feature importance analysis revealed both known and novel ECG correlates. ECGs hold promise as non-invasive, explainable biomarkers for selected neurocognitive disorders. This study demonstrates robust performance across cohorts and lays the groundwork for future clinical applications, including early detection and personalized monitoring.

\end{abstract}

\subsubsection*{What is already known on this topic}
Electrocardiogram abnormalities have been explored as potential biomarkers for neurocognitive disorders, but their diagnostic value has not been thoroughly investigated. Recent studies have only begun to examine these abnormalities in exploratory analyses to identify potential associations.

\subsubsection*{What this study adds}
This study demonstrates that ECG features can effectively predict neurocognitive disorders, extending beyond their traditional use as biomarkers for cardiac conditions. The findings show high predictive performance across both internal and external datasets. Additionally, the use of an explainable approach allows for a deeper understanding of the associations between specific ECG features and neurocognitive disorders, confirming existing knowledge while also providing new insights. This research sheds light on the physiological changes that may underlie the interplay between cardiac features and neurocognitive disorders.

\subsubsection*{How this study might affect research, practice or policy}
This work lays the foundation for integrating ECG into the diagnosis and monitoring of neurocognitive disorders. It presents a non-invasive, cost-effective tool that could aid as a companion tool in more efficient detection and personalized therapy management in clinical settings, with the potential to transform both research and clinical practices.

\section{Introduction}

\subsection{Clinical relevance}
Neurocognitive disorders are among the most challenging diseases, profoundly affecting both individuals and society. Disorders such as the underlaying pathology of dementia Alzheimer's disease \cite{karantzoulis2011distinguishing} and Parkinson's diseaselead to progressive cognitive and motor decline, severely limiting patients' independence and quality of life \cite{aarsland2021parkinson}. Other disorders like symphotmatic dementia and delirium further complicate mental health, often causing confusion, memory loss, and significant behavioral changes \cite{fong2022inter}. Notably, these disorders are closely interrelated, as they share common neurobiological pathways, genetic factors, and overlapping cognitive, affective, and behavioral symptoms. The distinction between them is often difficult, complicating diagnosis and treatment, and highlighting the need for integrated, multidisciplinary approaches in neurocognitive care \cite{taslim2024neuropsychiatric}. These disorders not only disrupt the lives of those affected but also impose a considerable burden on caregivers and healthcare systems \cite{eichel2022neuropsychiatric}. Early and fast diagnosis is crucial to effectively manage these diseases, but current diagnostic methods can be complex, costly, and sometimes inaccessible \cite{cummings2021role}.

\subsection{Emerging role of ECG beyond cardiology}
Electrocardiograms (ECG) have long been established as a critical tool in cardiology, used primarily to detect and monitor cardiac disorders by measuring the electrical activity of the heart. ECGs provide vital information about heart rhythms, helping to identify issues such as arrhythmias, ischemia, and other cardiac abnormalities \cite{siontis2021artificial}. While ECGs are traditionally associated with cardiac care, recent advancements in medical research have expanded their potential applications. Novel uses of ECG, supported by advancements in data analysis and machine learning, are now being explored in predictive modeling applications across fields beyond cardiology, including the estimation of laboratory values \cite{alcaraz2024cardiolab} and patient deterioration in emergency departments \cite{alcaraz2024mds}.  

\subsection{Emerging role of neurocardiology}
The potential link between ECG abnormalities and neurocognitive disorders is an emerging area of study. For instance, recent studies continue to highlight the complex interplay between mental and cardiovascular health, underscoring the clinical value of early detection and integrated care strategies. For instance, \cite{liu2025guidelines} discusses how depression can manifest with cardiovascular symptoms and proposes traditional interventions tailored to the underlying physiological patterns. \cite{ali2021guidelines} emphasizes the persistent gap between clinical guidelines and actual practice in monitoring cardiometabolic risk among antipsychotic users, where cardiovascular disease remains a leading cause of death. Additionally, \cite{li2024virtual} demonstrates that virtual reality-based psychological interventions can effectively reduce anxiety symptoms in acute myocardial infarction patients, offering an example of how digital solutions can be integrated into cardiovascular care. Conditions like Alzheimer's and Parkinson's disease may affect the autonomic nervous system \cite{gonccalves2022heart}, leading to detectable changes in heart rate variability and other ECG patterns. Similarly, psychiatric disorders such as dementia and delirium have been associated with cardiac function \cite{raberi2023postoperative}, potentially reflecting the underlying physiological disruptions.

\subsection{Current diagnostic practices}
To date, few studies have explored the diagnostic utility of ECG for neurocognitive disorders. A retrospective study linked specific ECG features to incident dementia and modestly improved risk prediction, but lacked machine learning to model complex patterns and did not validate findings across independent cohorts, limiting generalizability \cite{isaksen2022associations}. A study showed that ECG-derived heart rate variability could detect autonomic dysfunction in patients with REM sleep behavior disorder, an early sign of Parkinson’s disease but was limited by a small sample, lacked broader validation, and did not confirm progression to Parkinson’s \cite{valappil2010exploring}. Similarly, a recent study used machine learning to predict neurological outcomes in post-cardiac arrest patients; however, it relied on a single dataset, required EEG for strong performance since ECG-only models underperformed, and relied on synthetic data that may not reflect real-world populations \cite{niu2025explainable}.

\subsection{Contributions}
Based on these observations, this study investigates whether machine learning models trained on ECG data and basic demographic information can accurately identify specific neurocognitive disorders, namely dementia (including Alzheimer's disease and unspecified dementia), delirium, and Parkinson’s disease. We developed and validated models capable of distinguishing these conditions, demonstrating strong performance across both internal and external datasets. In addition, we provided transparent and explainable insights into the ECG features contributing to each diagnostic category. Our findings highlight the potential of ECG as a clinically useful, low-cost, and non-invasive diagnostic support tool in neurocognitive care, either as a standalone method or, more conservatively, as a complement to existing diagnostic practices. To our knowledge, this is the first study to apply explainable machine learning to ECG data for diagnosing these disorders, with validation across diverse clinical populations confirming its applicability.

\section{Methods}

\subsection{Dataset}

\begin{figure*}[!ht]
    \centering
    \includegraphics[width=\textwidth]{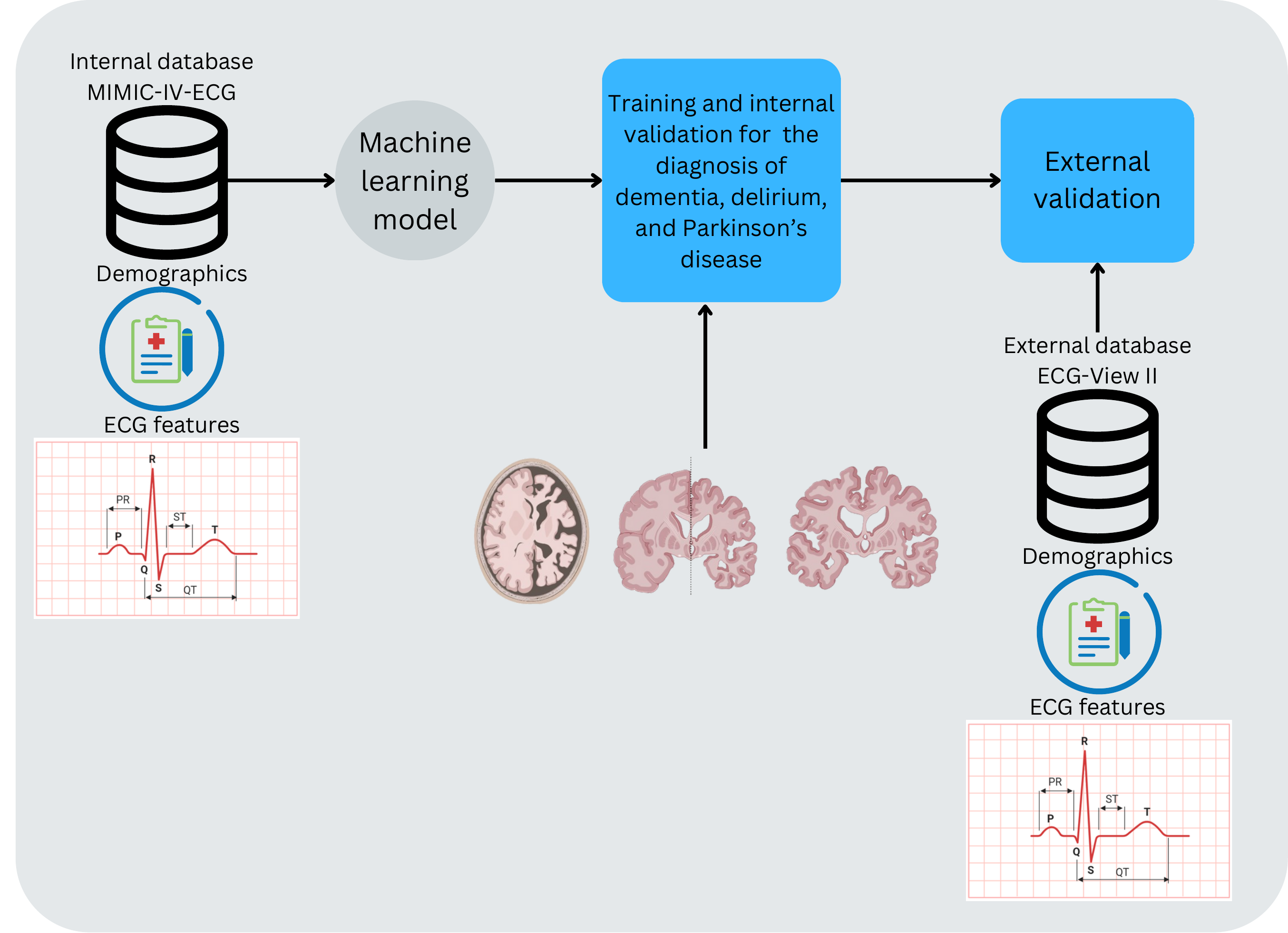}
    \caption{Diagrammatic illustration of our proposed methodology. The MIMIC-IV-ECG dataset serves as our internal dataset, providing demographic and ECG features used as input for training a tree-based model to diagnose various neurocognitive disorders. For external validation, we utilize a second patient cohort from the ECG-View II dataset, extracting the same set of features and neurocognitive targets. The disorders are defined based on ICD-10-CM codes.}
    \label{fig:abstract}
\end{figure*}

Figure~\ref{fig:abstract} contains a schematic representation of our proposed methodology in which the primary dataset for training and internal evaluation was obtained from the Medical Information Mart for Intensive Care (MIMIC)-IV-ECG database \cite{johnson2023mimic,MIMICIVECG2023}, a subset of a large-scale critical care dataset collected at the Beth Israel Deaconess Medical Center in Boston, Massachusetts. It includes data from patients admitted to the emergency department (ED) and intensive care unit (ICU). We included patients aged 18 years and older with at least one ECG record and corresponding discharge diagnoses recorded using ICD-10-CM codes between 2008 and 2019. MIMIC-IV-ECG is a publicly available, de-identified dataset, and thus individual patient consent was not required. The use of this dataset was approved by the institutional review boards of the Massachusetts Institute of Technology (MIT) and Beth Israel Deaconess Medical Center through Physionet.

We externally validate our findings using the ECG-VIEW-II database \cite{kim2017ecg}. ECG-VIEW-II includes data from patients at a South Korean tertiary teaching hospital. To create a comprehensive and harmonized feature set, ECG features from the MIMIC-IV dataset were aligned with those from the ECG-VIEW-II database where the standardized feature set consists of ECG-derived measurements (RR-interval, PR-interval, QRS-duration, QT-interval, QTc-interval in milliseconds; P-wave-axis, QRS-axis, and T-wave-axis in degrees), along with demographic attributes (binary sex and age as a continuous variable) which serves as our secondary dataset for external validation. Inclusion criteria for ECG-VIEW-II followed the same logic: adult patients (18+) with at least one ECG record and corresponding ICD-10-CM diagnosis between 1994 and 2015. ECG-VIEW-II is also a publicly available, de-identified dataset released under institutional review board approval, and thus patient consent was not applicable. Target variables are based on discharge diagnoses encoded using the International Classification of Diseases Clinical Modification (ICD-10-CM) \cite{WHO1992ICD10}, we primarly investigate the following disorders; G30: Alzheimer's disease, G20: Parkinson's disease, F01: Vascular dementia, F03: Dementia, and F05: Delirium (physiological).

For the internal dataset, stratified folds were created based on diagnoses, age, and gender distributions, using an 18:1:1 split as described in previous research \cite{strodthoff2024prospects}. The external evaluation is performed on the complete cohort. The training process emphasizes MIMIC-IV-ECG due to its broader ethnic diversity compared to ECG-VIEW-II, which helps improve the model’s generalization across various populations, as shown in prior studies \cite{alcaraz2024estimationcardiacnoncardiacdiagnosis}, which focused on cardiac disorders, \cite{alcaraz2024electrocardiogrambaseddiagnosisliverdiseases} for liver disorders, and \cite{alcaraz2024explainablemachinelearningneoplasms} for neoplasms. This strategy ensures robust internal training and reliable external validation across ethnically and geographically diverse cohorts.

\subsection{Models and evaluation}
In this study, we create individual tree-based models using Extreme Gradient Boosting (XGBoost) to tackle binary classification tasks, with a distinct model for each selected ICD-10-CM code. To avoid overfitting, we apply early stopping with a patience of 10 iterations on the validation fold during training. We did not apply any imputation into the dataset due to the inherently nature of the utilized model to handle missing data. Model performance is assessed using the area under the AUROC on the validation fold. We report AUROC scores for both the internal test set and an external dataset, along with 95\% confidence intervals calculated through empirical bootstrapping with 1000 iterations. To assess model calibration, we present calibration curves based on the internal test set for which we apply model-agnostic calibration and fit isotonic regression models on the validation set and report calibrated test set results. Additionally, we evaluate clinical utility using decision curve analysis, comparing our model's net benefit to standard strategies such as "refer all" and "refer none" \cite{vickers2006decision}.

\subsection{Explainability}
Our objective goes beyond merely assessing model performance. To gain deeper insights into the trained models, we integrate Shapley values into our workflow \cite{lundberg2020local}. These values provide a method for evaluating feature importance by measuring the individual contribution of each feature to the model's predictions.

\subsection{Study protocol and reporting standards}
This study follows a comprehensive protocol aligned with established guidelines, including TRIPOD (Transparent Reporting of a Multivariable Prediction Model for Individual Prognosis or Diagnosis), STARD (Standards for Reporting Diagnostic Accuracy Studies), and MI-CLAIM (Minimum information about clinical artificial intelligence modeling). Detailed documentation is provided in the supplementary material.

\section{Results}

\subsection{Descriptive statistics}

\begin{table}[!ht]
    \centering
    \begin{tabular}{lll}
    \hline
    \textbf{Variable} & \textbf{MIMIC-IV-ECG} & \textbf{ECG-View II} \\ \hline\hline
    \textbf{Gender (\%)} &  &  \\ \hline
    Female samples    & 226,892 (48.50)  & 375,733 (48.44) \\
    Male samples & 240,837 (51.49)  & 399,802 (51.55) \\ \hline
    \textbf{Age (\%)} &  &  \\ \hline
    Median years (IQR) & 66 (25) & 52 (25) \\ 
    Quantile 1 age range (\%) & 18-53 (23.83) & 18-40 (24.03) \\
    Quantile 2 age range (\%) & 53-66 (25.16) & 40-52 (25.75) \\
    Quantile 3 age range (\%) & 66-78 (25.60) & 52-65 (24.94) \\
    Quantile 4 age range (\%) & 78-101 (25.40) & 65-109 (25.28) \\ \hline
    \textbf{ECG features Median (IQR)} &  &  \\ \hline
    RR-interval[ms] & 769 (264)  & 857 (227) \\
    PR-interval[ms] & 158 (38)  & 158 (28) \\
    QRS-duration[ms] & 94 (23)  & 90 (14) \\
    QT-interval[ms] & 394 (68)  & 392 (48) \\
    QTc-interval[ms] & 447 (47)  & 421 (37) \\
    P-wave-axis[${}^\circ$] & 51 (32) & 53 (28) \\
    QRS-axis[${}^\circ$] & 13 (61) & 48 (49) \\
    T-wave-axis[${}^\circ$] & 42 (58)  & 44 (33) \\ \hline
    \end{tabular}%
    \caption{Descriptive statistics of the two ECG datasets, MIMIC-IV-ECG and ECG-View II. The table summarizes demographic variables gender (absolute sample count and percentage) and age (median and interquartile range) along with age group distributions by quantiles. ECG features are reported as median (IQR), with temporal intervals (RR, PR, QRS, QT, QTc) given in milliseconds (ms), and electrical axes (P-wave, QRS, T-wave) in degrees (${}^\circ$).}
    \label{tab:descriptive}
\end{table}

Table~\ref{tab:descriptive} highlights key demographic and ECG feature differences between MIMIC-IV-ECG and ECG-View II. Both datasets have a nearly identical gender distribution with slightly more male than female samples. However, MIMIC-IV-ECG includes an older population (median age: 66 vs. 52 years), which may influence ECG characteristics. Notable differences include a shorter RR-interval in MIMIC-IV-ECG (769 ms vs. 857 ms) and a higher QTc-interval (447 ms vs. 421 ms), suggesting potential variations in heart rate and repolarization patterns. Additionally, the QRS axis shows a substantial shift (13° vs. 48°), which may reflect the differences in patient populations. A detailed comparison of ECG features between diagnostic-positive and -negative groups across the MIMIC-IV and ECG-VIEW II datasets is presented in the supplementary Table~\ref{tab:comparison}. The analysis reveals modest yet consistent feature differences, suggesting potential disease-related alterations in ECG signals in an exploratory manner.

\subsection{Discriminative, calibration, decision, and interpretability analysis}

\begin{figure*}[!ht]
    \centering
    \includegraphics[width=\textwidth]{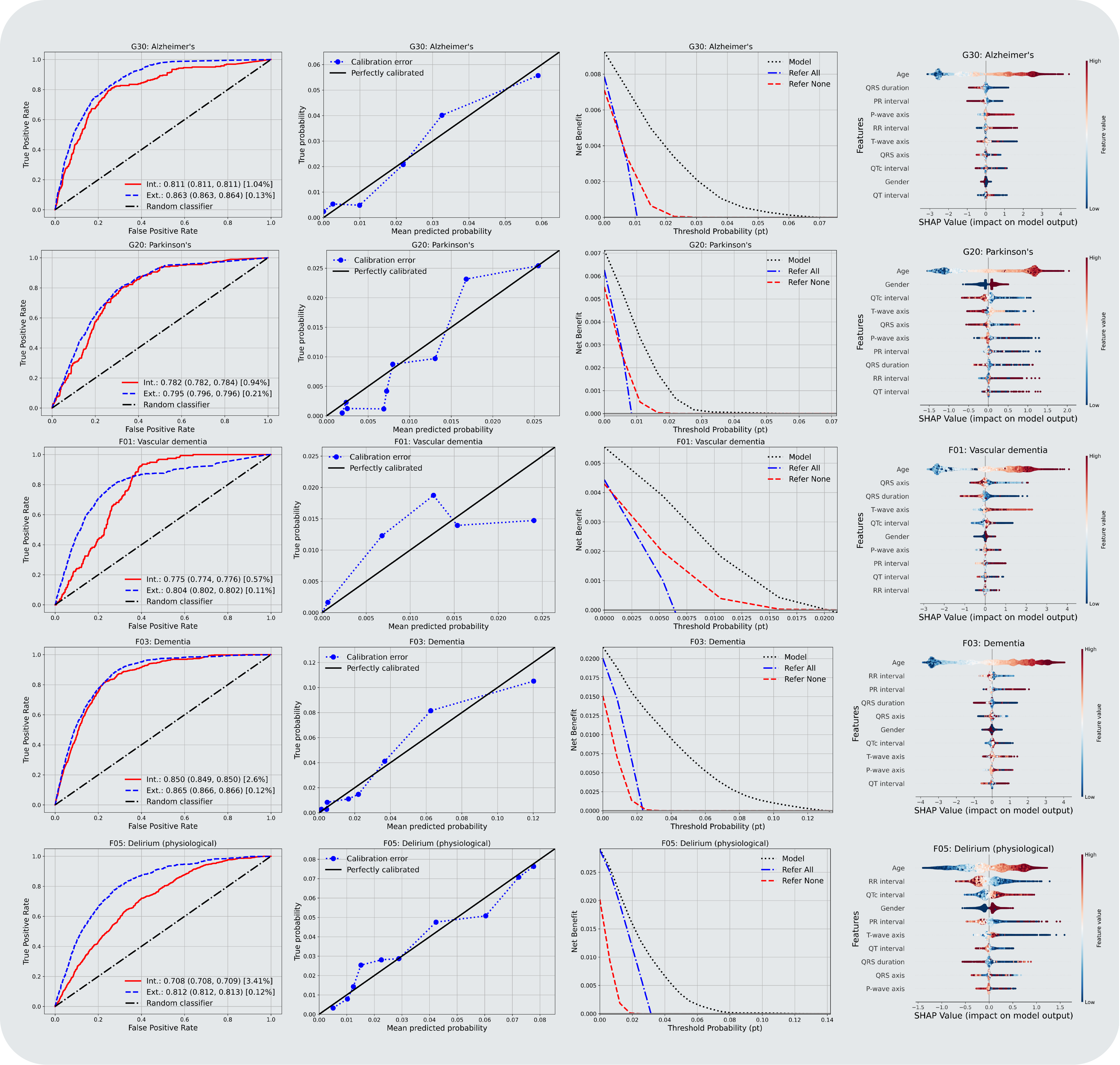}
    \caption{Model evaluation results across five neurocognitive disorders. Columns represent: (1) Discriminative performance (AUROC with 95\% CI), (2) Calibration plots, (3) Net benefit decision curves, and (4) SHAP-based feature importance. Rows correspond to the following ICD-10 codes; G30: Alzheimer's disease, G20: Parkinson's disease, F01: Vascular dementia, F03: Unspecified dementia, and F05: Delirium due to physiological condition.}
    \label{fig:results}
\end{figure*}

Comprehensively outlined by \cite{vancalster2024performanceevaluationpredictiveai}, assessing the predictive performance of AI algorithms in medicine has to cover several different categories, most notably, discrimination, clinical utility. We assess discimination in terms of receiver operating curves/AUROC scores, calibration through calibration plots and clinical utility through decision curve analysis/net benefit \cite{vickers2006decision}. This is complemented through an explainability analysis by means of SHAP values. We showcase each of the four categories as columns in Figure~\ref{fig:results}.

The first column in Figure~\ref{fig:results} presents the predictive performance of our model for various neurocognitive disorders, evaluated using AUROC scores on both internal (MIMIC-IV) and external (ECG-View) test sets. The 95\% prediction intervals offer insight into the reliability of these scores. Additionally, the class prevalence for each condition within its respective dataset is provided, offering context on the population distribution and demostrating predictive performance robustness due to their prevalence missmatch. In the MIMIC-IV cohort, prevalences range from 0.5\% to 3.41\%, while the ECG-View cohort exhibits much lower prevalences, from 0.05\% to 0.21\%. Significantly, the disorders with the highest predictive performance in the is F03: Dementia, with an internal AUROC of 0.848 (95\% CI: 0.848–0.848) and an external AUROC of 0.865 (0.864-0.965), followed by G30: Alzheimer's, with an internal AUROC of 0.809 (95\% CI: 0.808–0.810) and an external AUROC of 0.863 (95\% CI: 0.863–0.864). These findings highlight the reliability of our model in predicting neurocognitive disorders, despite differences in prevalence across datasets.

The second column in Figure~\ref{fig:results} illustrates the calibration of the model predictions for each condition, comparing predicted probabilities with observed outcomes. Across all conditions, our models demonstrate good calibration, indicating that the predicted probabilities are consistent with actual event rates. This is particularly relevant in clinical applications, where over- or underestimation of risk could have critical consequences. The reliability of these probability estimates further supports the model’s utility in risk stratification and decision support.

The third column in Figure~\ref{fig:results} presents decision curve analyses, assessing the clinical utility of our models. For all examined disorders, the models provide a higher net benefit across a range of threshold probabilities compared to the 'refer-all' and 'refer-none' strategies. This suggests that using our models to guide referral decisions could lead to better patient outcomes while minimizing unnecessary evaluations. The consistent superiority of our models in net benefit reinforces their practical value in aiding early detection and triage for neurocognitive disorders.

The fourth column in Figure~\ref{fig:results} represents the result of the interpretability analysis. Across all disorders investigated, age is the most important predictive feature, where increased age contributes positively. For the investigated neurocognitive disorders, QTc interval is the most important ECG feature for Parkinson's, where low values and high values of the marker contribute the most positively to each condition respectively. Similarly, T-wave axis is the second most important ECG feature for the same two disorders, where low values of the marker contributes the most positively to each condition. For the investigated psychiatric disorders, RR-interval is the most important predictive ECG feature across unspecified dementia and delirium due to known physiological disorders, where low values of the marker mainly contribute to the disorders. Similarly, PR-interval is the second most important ECG feature for the same two disorders, where mainly high and low values of the marker contribute positively to the disorders.

Finally, we investigate ICD-10 codes with strong internal discriminative performance on the MIMIC-IV-ECG dataset that could not be extermally evaluated on the ECG-View II. Some of the high-performing categories include eating disorders such as F502: Bulimia nervosa (AUROC = 0.953) and F509: Eating disorder, unspecified (0.916), personality disorders such as F602: Antisocial personality disorder (0.952), and conduct disorders such as F91: Conduct disorders (0.878). Additionally, several substance-related disorders showed strong internal performance, including F1510: Other stimulant abuse, uncomplicated (0.903), F1220: Cannabis dependence, uncomplicated (0.891), F1310: Sedative, hypnotic or anxiolytic abuse, uncomplicated (0.884), F1910: Other psychoactive substance abuse, uncomplicated (0.895), and F1012: Alcohol abuse with intoxication (0.873). The full list of these codes is provided in the supplementary Table~\ref{tab:noext}, including 8 codes with AUROCs above 0.9 and 40 codes between 0.8 and 0.9, representing promising directions for future investigation. Additionally, the supplementary Table~\ref{tab:poorext} lists conditions that showed good performance on the internal test set but failed to generalize externally. For instance, G931 (Anoxic brain damage) was excluded from the final analysis despite strong internal and external AUROC values, due to poor calibration and limited clinical utility (see the supplementary Figure~\ref{fig:anoxic}).

\section{Discussion}

\subsection{Main findings}
Detecting neurocognitive disorders through ECG features may initially seem unconventional, as the ECG is traditionally associated with cardiovascular diagnoses. However, the physiological interplay between the cardiovascular and neurocognitive systems offers a unique opportunity for diagnostic innovation. While the mechanisms linking neurocognitive disorders to ECG abnormalities remain incompletely understood, some known relations include autonomic nervous system dysfunction, neurocognitive changes, vascular alterations, electrophysiological disruptions, and hormonal or metabolic factors. For example, Alzheimer's disease, Parkinson's disease, and brain damage share common ECG abnormalities, primarily reflecting autonomic dysfunction and brain-heart communication. 

Across the investigated disorders, reduced heart rate variability (HRV), bradycardia due to parasympathetic dominance, and arrhythmias like atrial fibrillation \cite{barkhordarian2024atrial} are frequently observed. Similarly, medication effects, such as those from cholinesterase inhibitors in Alzheimer’s or dopamine agonists in Parkinson’s, can further influence conduction and structural heart changes \cite{huang2020comparative}. Despite these insights, the specific relationship to certain abnormalities remains unclear. Our findings reveal specific ECG patterns as distinctive features for these disorders, identified using machine learning. Similarly, beyond diagnosis, ECG features hold promise for therapy management by monitoring treatment responses, risk stratification for medication-related adverse events, early intervention to identify stress or relapse, and longitudinal monitoring for chronic disorders. This interdisciplinary approach bridges neurology, psychiatry, and cardiology, advancing novel diagnostic and therapeutic strategies.

The exceptional predictive power of specific ECG features highlights their ability to reliably identify neurocognitive disorders from a single ECG. Consistently high AUROC scores in both internal and external validations demonstrate the robustness of these features across diverse cohorts. The distinct patterns observed across physiological systems underscore the intricate connection between cardiac and neurocognitive health. In this study, age was identified as a key factor, with older patients contributing more to most disorders. This is consistent with well-documented evidence of brain and structural degeneration associated with aging. For example, previous research has reported abnormal electroencephalograms in patients over 45 years old with disorders such as senile and arteriosclerotic psychosis, involutional psychosis, psychosis with mental deficiency, manic-depressive states, psychoneurosis, and schizophrenia \cite{doi:10.1176/ajp.101.1.82}.

A recent study investigated the association of ECG features and Alzheimer's disease and have found high hazard ratios for low QRS duration and for low PR interval \cite{isaksen2023association}, which aligns with our findings. Another study reported abormally high P wave axes\cite{gutierrez2019association}, which also aligns with our findings. For Parkinson's disease, a study that collected data from various articles and meta-analyses found a clear preponderance of higher incidence and prevalence in males \cite{georgiev2017gender}, which also aligns with our findings. Similarly, a study conducted on 981 patients demonstrated that high QT interval values are associated with Parkinson's disease, which again matches our findings. However, the authors report that high QTc interval values were also observed in their cohort \cite{rabkin2024qt}. In contrast, our analysis, based on nearly 4,000 samples, suggests that while the raw QT interval is elevated in Parkinson's patients, the QTc interval, when corrected for heart rate, appears to be lower. This discrepancy may be due to differences in sample size, patient characteristics, or the methodologies used to correct the QT interval, as well as the larger heterogeneity within our cohort. Regarding all-cause, left QRS axis deviation have been associated with patients by a recent study \cite{mao2023ventricular}, which aligns with low or negative values found in our findings. A similar study correlates with another of our findings, as the authors found that shorter QRS duration is associated with non-Alzheimer dementia \cite{isaksen2022associations}. For vascular dementia, prior work identified a significant increase of QTc values on a population group which also match our findings \cite{matei2015qt}. Finally, the high values of QTc interval for patients with delirium correlates with previous literature work on patients with delirium tremens whose also developed tachyarrhythmias and returned to sinus rhythm after appropriate treatment.

From the conditions with high predictive internal performance that can not be externally validated due to missing labels in the external dataset, the high discriminative performance of certain psychiatric and behavioral diagnoses such as eating disorders, substance use disorders, and personality disorders raises important clinical considerations about the physiological manifestations of these conditions. Traditionally, the diagnosis of such disorders relies on behavioral assessments, structured interviews, and, in cases of substance use, confirmatory tests like toxicology screenings. However, the ability of ECG-based models to predict these diagnoses suggests that these conditions may also leave detectable and specific cardi-physiological footprints, possibly through chronic stress, autonomic dysfunction, or cardiotoxic effects of certain substances.

The results presented in this work suggests on potentially new ECG features for neurocognitive disorders. This includes high values of RR interval for Alzheimer's disease, low values of QRS axis for Parkinson's disease, and low values of QRS duration for vascular dementia. We hypothesize that the high RR interval in Alzheimer's may reflect altered autonomic regulation due to parasympathetic dysfunction \cite{femminella2014autonomic}. The low QRS axis in Parkinson's disease could be linked to basal ganglia dysfunction affecting autonomic control of the heart \cite{zesiewicz2003autonomic}. Similarly, the low QRS duration in vascular dementia might result from cerebral ischemia and microvascular changes affecting cardiac conduction \cite{wang2018dysfunction}. These hypotheses warrant further investigation.

\subsection{Limitations}

ECG is a valuable tool for identifying electrical abnormalities that may correlate with neurocognitive disorders; however, it does not provide direct diagnostic confirmation or insights into specific neurophysiological changes underlying these disorders. Many ECG alterations are non-specific and can be attributed to factors such as stress, medication effects, or comorbid disorders, making it challenging to isolate patterns uniquely associated with neurocognitive disorders. Additionally, the causal mechanisms linking ECG abnormalities to neurocognitive disorders remain poorly understood, highlighting the need for further investigation into these complex relationships. Nevertheless, we want to raise that previous work also in ECG for diagnoses via ICD-10 in MIMIC-IV has shown that there is none significant label correlations \cite{strodthoff2024prospects}, which can be used as a suggestion against confounding effects through co-occurring diseases. Similarly, the work in \cite{isaksen2022associations} also suggests the feasibility of finding ECG features abnormalities on incident dementia patients (without treatments) furthing supporting the value of this work.

Future research should explore how ECG abnormalities differ across diverse demographic groups and distinguish them from normal variations, such as those related to age \cite{ott2024using}. Investigating the causal connections between ECG features and neurocognitive disorders will be essential for advancing our understanding \cite{alcaraz2024causalconceptts}. Further studies should focus on using raw ECG waveforms and validating findings across external datasets to improve diagnostic accuracy \cite{strodthoff2024prospects, alcaraz2024mds}. The demonstrated potential of raw ECG waveforms to outperform traditional ECG features emphasizes the importance of refining these approaches to enhance precision and reliability in detecting neurocognitive disorders.

\subsection{Implications}

Our approach naturally encompasses both therapy-naive patients and those attending follow-up visits, as ICD-10 codes represent a combination of newly assigned diagnoses and ongoing treatment cases. This creates potential confounding effects, where predictions might reflect therapy-related cardiotoxic changes rather than the neurocognitive condition itself. Notably, medications used for the treatment of Alzheimer’s disease are not completely safe in terms of cardiac side effects. Acetylcholinesterase inhibitors (Donepezil, Rivastigmine, Galantamine) can, in rare cases, cause bradycardia or QT interval prolongation, particularly in patients with preexisting cardiac arrhythmias \cite{howes2014cardiovascular}. In contrast, Memantine has minimal cardiovascular side effects. Additionally, some Parkinson’s disease medications have been associated with cardiac complications. Certain dopamine agonists, particularly ergot-derived agents such as pergolide and cabergoline, can increase the risk of valvular heart disease by stimulating serotonin 5-HT$_{2}B$ receptors, leading to fibrotic changes in heart valves \cite{schade2007dopamine}. In addition, antipsychotic medication is commonly used in neurocognitive disorders, which are associated with tachycardia, bradycardia and QTc prolongation \cite{buckley2000cardiovascular}

These therapy-related cardiac effects further complicate the distinction between disease progression and treatment influence, underscoring the need for careful monitoring, especially in patients with known cardiac disorders \cite{howes2014cardiovascular}. Future research should aim to stratify newly diagnosed cases from follow-ups, as such differentiation would enable a more precise evaluation of the model’s ability to distinguish between new diagnoses and therapy-induced patterns.

ECG innovations for neurocognitive disorders hold great promise across diagnosis, therapy management, and personalized care. Distinctive ECG patterns can act as features for more efficient, non-invasive detection of neurocognitive disorders. Beyond diagnosis, ECG monitoring contributes to therapy management by tracking treatment responses and identifying early physiological changes related to medication efficacy or adverse effects. Risk stratification also emerges as a vital application, enabling clinicians to predict adverse events such as arrhythmias linked to neurocognitive treatments or stress-related cardiovascular complications.

Longitudinal monitoring through periodic or wearable ECG devices provides valuable insights into disease progression, relapse or episodes prediction, facilitating real-time assessments of mental and physical health. ECG-derived features further support personalized medicine by helping tailor treatments to individual patients. Additionally, early intervention strategies can use subtle ECG changes to detect stress, anxiety, or relapse indicators, ensuring timely clinical responses, see a recent perspective emphasizes that early detection of cognitive impairment in primary care enables earlier interventions, including lifestyle changes and management of comorbid conditions, that can slow cognitive decline and improve overall outcomes for patients and caregivers \cite{fowler2025implementing}. Similarly, Recent work highlights the importance of predictive models for neurocognitive disorders, showing that early detection and intervention significantly enhance patient outcomes by enabling timely, tailored therapeutic strategies \cite{tsiakiri2024predictive}.

\section{Statements and declarations}

\subsection*{Contributorship}
JMLA, WH, and NS conceived and designed the project. JMLA conducted the full experimental analyses, with WH, and NS supervising them, and WH, EO, and DT providing critical revision of clinical intellectual content. JMLA produced the first draft, the rest of the authors revised it. All authors critically revised the content and approved the final version for publication. Computing resources were provided through Carl von Ossietzky University of Oldenburg.

\subsection*{Funding statement}
No specific funding was received for this research.

\subsection*{Competing interests} 
The authors have declared no competing interests.

\subsection*{Data and code availability}
The datasets can be found in raw format in their appropiate source codes \url{https://physionet.org/content/mimic-iv-ecg/1.0/}, and \url{http://ecgview.org/}. The code for dataset preprocessing and reproducing experiments is available at \url{https://github.com/AI4HealthUOL/CardioDiag}.

\subsection*{Ethical statement} 
This study used publicly available and de-identified datasets. No direct patient interaction or intervention was involved. As these datasets are released under established data use agreements and have been ethically approved for secondary research, additional ethics approval was not sought.

\bibliography{refs}

\clearpage

\renewcommand{\thetable}{\arabic{table}}
\setcounter{table}{0}

\renewcommand{\thefigure}{\arabic{figure}}
\setcounter{figure}{0}

\appendix

\section{Results for additional conditions}
\begin{figure}
    \centering
    \includegraphics[width=1\linewidth]{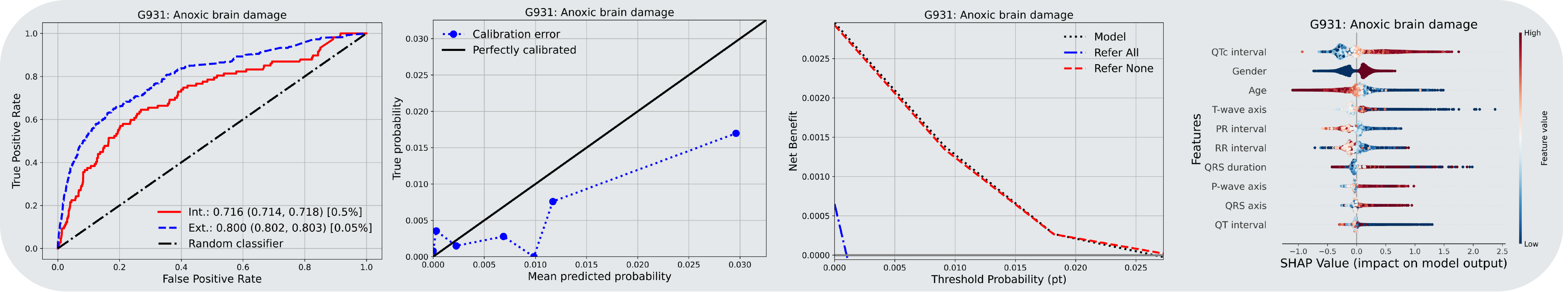}
    \caption{Performance in terms of AUROC, calibration curve, net benefit, as well as explainability for G931: Anoxic brain damage.}
    \label{fig:anoxic}
\end{figure}

In Table~\ref{tab:noext}, we show conditions with strong predictive performance on the internal dataset that could not be externally validated due to a lack of corresponding labels on the external dataset. In Table~\ref{tab:poorext}, we list conditions with strong internal predictive performance that showed a poor performance on the external dataset. In Figure~\ref{fig:anoxic}, we show the performance results for the condition G931: Anoxic brain damage.

\begin{longtable}{l|c}
    \hline
        \textbf{ICD-10 Code} \textbf{Description} & \textbf{AUROC (Internal)}  \\
        \hline
        F5001: norexia nervosa, restricting type. & 0.994 \\
F500: Anorexia nervosa. & 0.990  \\
F5000: Anorexia nervosa, unspecified. & 0.987 \\ 
F50: Eating disorders. & 0.962 \\
F502: Bulimia nervosa. & 0.953  \\
F602: Antisocial personality disorder. & 0.952 \\ 
F509: Eating disorder, unspecified. & 0.916  \\ 
F1510: Other stimulant abuse, uncomplicated. & 0.903 \\ 
        \hline
F1910: Other psychoactive substance abuse, uncomplicated. & 0.895 \\
F1220: Cannabis dependence, uncomplicated. & 0.891 \\
F151: Other stimulant abuse. & 0.886 \\
F1310: Sedative, hypnotic or anxiolytic abuse, uncomplicated. & 0.884  \\
F12: Cannabis related disorders. & 0.883 \\
F121: Cannabis abuse. & 0.879 \\
F91: Conduct disorders. & 0.878 \\ 
F1012: Alcohol abuse with intoxication. & 0.873 \\
F149: Cocaine use, unspecified. & 0.871 \\
F1210: Cannabis abuse, uncomplicated. & 0.870 \\
F13: Sedative, hypnotic, or anxiolytic related disorders. & 0.869  \\
F15: Other stimulant related disorders. & 0.868  \\
F1023: Alcohol dependence with withdrawal. & 0.860  \\
F039: Unspecified dementia. & 0.849  \\
F1290: Cannabis use, unspecified, uncomplicated. & 0.846 \\
F129: Cannabis use, unspecified. & 0.845 \\
F1022: Alcohol dependence with intoxication. & 0.845 \\
F0390: Unspecified dementia, unspecified severity... & 0.843  \\
F1490: Cocaine use, unspecified, uncomplicated. & 0.843  \\
F603: Borderline personality disorder. & 0.840 \\
F250: Schizoaffective disorder, bipolar type. & 0.834 \\
F068: Other specified mental disorders due to known physiological condition. & 0.834 \\
F11: Opioid related disorders. & 0.833 \\
F132: Sedative, hypnotic or anxiolytic-related dependence. & 0.833  \\
F1123: Opioid dependence with withdrawal. & 0.829 \\
F112: Opioid dependence. & 0.823 \\
F0391: Unspecified dementia with behavioral disturbance. & 0.821 \\
F111: Opioid abuse. & 0.820 \\
F0280: Dementia in other diseases classified elsewhere, unspecified severity... & 0.820  \\
F1120: Opioid dependence, uncomplicated. & 0.819 \\
F141: Cocaine abuse. & 0.818 \\
F1110: Opioid abuse, uncomplicated. & 0.816 \\
F1993: Other psychoactive substance use, unspecified with withdrawal. & 0.815 \\
F102: Alcohol dependence. & 0.811  \\
F1920: Other psychoactive substance dependence, uncomplicated. & 0.810 \\
F1190: Opioid use, unspecified, uncomplicated. & 0.808 \\
F14: Cocaine related disorders. & 0.803 \\
F181: Inhalant abuse. & 0.801 \\
F1810: Inhalant abuse, uncomplicated. & 0.801 \\
F1410: Cocaine abuse, uncomplicated. & 0.800 \\
F18: Inhalant related disorders. & 0.800 \\
        \hline
F909: Attention-deficit hyperactivity disorder, unspecified type. & 0.797 \\
F1020: Alcohol dependence, uncomplicated. & 0.795 \\
F98: Other behavioral and emotional disorders... & 0.795 \\
F02: Dementia in other diseases classified elsewhere. & 0.789 \\
F028: Dementia in other diseases classified elsewhere. & 0.789  \\
F205: Residual schizophrenia. & 0.788 \\
F192: Other psychoactive substance dependence. & 0.787 \\
F90: Attention-deficit hyperactivity disorders. & 0.783 \\
F10: Alcohol related disorders. & 0.782 \\
F0150: Vascular dementia, unspecified severity, without disturbance. & 0.777 \\
F1320: Sedative, hypnotic or anxiolytic dependence, uncomplicated. & 0.775 \\
F0151: Vascular dementia with behavioral disturbance. & 0.773  \\
F1420: Cocaine dependence, uncomplicated. & 0.767 \\
F4322: Adjustment disorder with anxiety. & 0.767 \\
F131: Sedative, hypnotic or anxiolytic-related abuse. & 0.754 \\
F609: Personality disorder, unspecified. & 0.752 \\
F60: Specific personality disorders. & 0.751 \\
F259: Schizoaffective disorder, unspecified. & 0.749 \\
F1121: Opioid dependence, in remission. & 0.745 \\
F101: Alcohol abuse. & 0.743 \\
F119: Opioid use, unspecified. & 0.743 \\
F39: Unspecified mood [affective] disorder. & 0.734 \\
F25: Schizoaffective disorders. & 0.732 \\
F429: Obsessive-compulsive disorder, unspecified. & 0.732 \\
F122: Cannabis dependence. & 0.729 \\
F1011: Alcohol abuse, in remission. & 0.727 \\
F1010: Alcohol abuse, uncomplicated. & 0.725 \\
F142: Cocaine dependence. & 0.723 \\
F191: Other psychoactive substance abuse. & 0.720 \\
F42: Obsessive-compulsive disorder. & 0.720 \\
F109: Alcohol use, unspecified. & 0.712 \\
F1721: Nicotine dependence, cigarettes. & 0.710 \\
F332: Major depressive disorder, recurrent severe without psychotic markers. & 0.702 \\
F1021: Alcohol dependence, in remission. & 0.701 \\ \hline
\caption{ICD-10 codes of neuropsychiatric disorders with high predictive performance in the internal dataset that are not present in the external dataset and can therefore not be externally validated}
\label{tab:noext}
\end{longtable}

\begin{table}[ht]
    \centering
\small
    \begin{tabular}{l|c|c}
    \hline
        \textbf{ICD-10 Code} \textbf{Description} & \textbf{AUROC (Internal)} & \textbf{AUROC (External)}  \\
        \hline
F311: Bipolar disorder, current episode manic without psychotic markers. & 0.715  & 0.628  \\
F322: Major depressive disorder, single episode, severe without psychotic markers. & 0.7956 & 0.520 \\ 
G931: Anoxic brain damage & 0.717 & 0.800 \\ \hline
    \end{tabular}
    \caption{ICD-10 codes of neuropsychiatric disorders with internal AUROC above 0.7 that does not have at least 0.7 in external or does not fall into the neurocognitive disorders category}
    \label{tab:poorext}
\end{table}

\section{Feature comparison across binary outcomes}

\begin{table*}[ht!]
    \centering
    \small
    \resizebox{\textwidth}{!}{%
    \begin{tabular}{llcccc}
    \hline
    \textbf{ICD-10 Code: Description} & \textbf{Feature} & \textbf{MIMIC-IV Negative} & \textbf{MIMIC-IV Positive} & \textbf{ECG-VIEW II Negative} & \textbf{ECG-VIEW II Positive} \\
    \hline\hline

    \multirow{8}{*}{G30: Alzheimer's disease}
& RR & 769.00 (645.00, 909.00) & 769.00 (638.00, 909.00) & 857.00 (741.00, 968.00) & 870.00 (750.00, 984.00) \\
& PR $\uparrow$ & 158.00 (140.00, 178.00) & 162.00 (142.00, 185.00) & 158.00 (144.00, 172.00) & 162.00 (148.00, 180.00) \\
& QRS & 94.00 (85.00, 107.00) & 94.00 (84.00, 114.00) & 90.00 (84.00, 98.00) & 88.00 (80.00, 96.00) \\
& QT $\uparrow$ & 394.00 (360.00, 428.00) & 400.00 (364.00, 436.00) & 392.00 (368.00, 416.00) & 404.00 (378.00, 430.00) \\
& QTc $\uparrow$ & 447.44 (426.03, 473.33) & 455.69 (434.61, 482.02) & 421.00 (405.00, 442.00) & 435.00 (417.00, 454.00) \\
& P wave axis $\uparrow$ & 51.00 (32.00, 64.00) & 52.50 (32.00, 67.00) & 53.00 (37.00, 65.00) & 54.00 (38.00, 66.00) \\
& QRS axis $\downarrow$ & 14.00 (-15.00, 46.00) & 0.00 (-31.00, 31.00) & 48.00 (19.00, 68.00) & 36.00 (10.00, 55.00) \\
& T wave axis $\uparrow$ & 42.00 (13.00, 71.00) & 48.00 (12.00, 81.00) & 44.00 (27.00, 60.00) & 50.00 (32.00, 64.00) \\
    \hline

    \multirow{8}{*}{G20: Parkinson's disease}
& RR & 769.00 (645.00, 909.00) & 800.00 (667.00, 923.00) & 857.00 (741.00, 968.00) & 800.00 (690.00, 909.00) \\
& PR $\uparrow$ & 158.00 (140.00, 178.00) & 164.00 (144.00, 187.00) & 158.00 (144.00, 172.00) & 160.00 (146.00, 178.00) \\
& QRS & 94.00 (85.00, 107.00) & 96.00 (86.00, 113.00) & 90.00 (84.00, 98.00) & 86.00 (80.00, 96.00) \\
& QT & 394.00 (360.00, 428.00) & 400.00 (366.00, 434.00) & 392.00 (368.00, 416.00) & 390.00 (364.00, 416.00) \\
& QTc $\uparrow$ & 447.65 (426.08, 473.38) & 447.82 (426.56, 474.31) & 421.00 (405.00, 442.00) & 433.00 (411.00, 457.00) \\
& P wave axis $\downarrow$ & 51.00 (32.00, 64.00) & 48.00 (26.00, 64.00) & 53.00 (37.00, 65.00) & 52.00 (34.00, 67.00) \\
& QRS axis $\downarrow$ & 14.00 (-15.00, 46.00) & 0.00 (-30.00, 29.00) & 48.00 (19.00, 68.00) & 25.00 (-2.00, 51.00) \\
& T wave axis $\uparrow$ & 42.00 (13.00, 71.00) & 48.00 (15.00, 77.00) & 44.00 (27.00, 60.00) & 49.00 (29.00, 66.00) \\
    \hline
    \multirow{8}{*}{F01: Vascular dementia}
& RR & 769.00 (645.00, 909.00) & 769.00 (652.00, 909.00) & 857.00 (741.00, 968.00) & 779.00 (667.00, 909.00) \\
& PR $\uparrow$ & 158.00 (140.00, 178.00) & 164.00 (144.00, 188.00) & 158.00 (144.00, 172.00) & 164.00 (146.00, 182.00) \\
& QRS & 94.00 (85.00, 107.00) & 96.00 (86.00, 120.00) & 90.00 (84.00, 98.00) & 88.00 (82.00, 98.00) \\
& QT $\uparrow$ & 394.00 (360.00, 428.00) & 404.00 (368.00, 439.00) & 392.00 (368.00, 416.00) & 396.00 (366.00, 428.00) \\
& QTc $\uparrow$ & 447.44 (426.04, 473.33) & 460.13 (437.80, 484.80) & 421.00 (405.00, 442.00) & 446.00 (426.00, 467.00) \\
& P wave axis & 51.00 (32.00, 64.00) & 51.00 (27.50, 65.00) & 53.00 (37.00, 65.00) & 51.00 (33.00, 64.00) \\
& QRS axis $\downarrow$ & 14.00 (-15.00, 46.00) & 0.00 (-33.00, 32.00) & 48.00 (19.00, 68.00) & 33.00 (5.00, 60.00) \\
& T wave axis $\uparrow$ & 42.00 (13.00, 71.00) & 54.00 (15.00, 95.00) & 44.00 (27.00, 60.00) & 53.00 (29.00, 78.00) \\
    \hline

    \multirow{8}{*}{F03: Dementia}
& RR $\downarrow$ & 769.00 (645.00, 909.00) & 759.00 (638.00, 909.00) & 857.00 (741.00, 968.00) & 800.00 (682.00, 938.00) \\
& PR $\uparrow$ & 158.00 (140.00, 178.00) & 164.00 (142.00, 188.00) & 158.00 (144.00, 172.00) & 160.00 (144.00, 176.00) \\
& QRS & 94.00 (85.00, 107.00) & 97.00 (86.00, 121.00) & 90.00 (84.00, 98.00) & 86.00 (80.00, 94.00) \\
& QT $\uparrow$ & 394.00 (360.00, 428.00) & 400.00 (364.00, 437.00) & 392.00 (368.00, 416.00) & 394.00 (366.00, 424.00) \\
& QTc $\uparrow$ & 447.21 (426.01, 473.12) & 459.20 (436.36, 485.76) & 421.00 (405.00, 442.00) & 437.00 (415.00, 460.00) \\
& P wave axis & 51.00 (32.00, 64.00) & 51.00 (27.00, 67.00) & 53.00 (37.00, 65.00) & 52.00 (35.00, 65.00) \\
& QRS axis $\downarrow$ & 14.00 (-14.00, 47.00) & -2.00 (-32.00, 30.00) & 48.00 (19.00, 68.00) & 27.00 (3.00, 54.00) \\
& T wave axis & 42.00 (14.00, 71.00) & 50.00 (12.00, 87.00) & 44.00 (27.00, 60.00) & 44.00 (24.75, 67.00) \\
    \hline

    \multirow{8}{*}{F05: Delirium (physiological)}
& RR $\downarrow$ & 769.00 (645.00, 909.00) & 714.00 (594.00, 845.00) & 857.00 (741.00, 968.00) & 723.00 (600.00, 857.00) \\
& PR $\downarrow$ & 158.00 (140.00, 178.00) & 154.00 (136.00, 178.00) & 158.00 (144.00, 172.00) & 156.00 (140.00, 174.00) \\
& QRS & 94.00 (85.00, 107.00) & 96.00 (86.00, 116.00) & 90.00 (84.00, 98.00) & 88.00 (80.00, 98.00) \\
& QT $\downarrow$ & 394.00 (360.00, 428.00) & 386.00 (348.00, 424.00) & 392.00 (368.00, 416.00) & 380.00 (348.00, 416.00) \\
& QTc $\uparrow$ & 447.21 (426.00, 473.08) & 458.39 (435.04, 485.35) & 421.00 (405.00, 442.00) & 448.00 (422.00, 472.00) \\
& P wave axis & 51.00 (32.00, 64.00) & 52.00 (31.00, 66.00) & 53.00 (37.00, 65.00) & 52.00 (36.00, 67.00) \\
& QRS axis $\downarrow$ & 14.00 (-15.00, 46.00) & 6.00 (-24.00, 41.00) & 48.00 (19.00, 68.00) & 35.00 (4.00, 63.00) \\
& T wave axis $\uparrow$ & 42.00 (14.00, 70.00) & 48.00 (8.00, 85.00) & 44.00 (27.00, 60.00) & 57.00 (31.00, 87.50) \\
    \hline
    \end{tabular}%
    }
    \caption{Comparison of ECG-derived feature medians and interquantile ranges (IQR) between patients with and without neurocognitive diagnoses in the MIMIC-IV and ECG-VIEW II cohorts. Neurocognitive disorders analyzed include Alzheimer's disease (G30), and Parkinson's disease (G20), vascular dementia (F01), unspecified dementia (F03), and delirium due to physiological disorders (F05). Feature distributions are reported separately for positive (with diagnosis) and negative (without diagnosis) cases in each cohort. To highlight consistent trends across both cohorts, an uparrow $\uparrow$ indicates features with increased values in positive cases, while $\downarrow$ indicates features with decreased values}
    \label{tab:comparison}
\end{table*}

Table~\ref{tab:comparison} presents median values with interquartile ranges for various ECG features stratified by diagnosis codes related to neurocognitive disorders across two datasets. For each ICD-10 code, values are compared between patients without (negative) and with (positive) diagnosis. Notably, for Alzheimer's disease, the positive samples against the negative ones show an increase in PR, QTc, P wave axis, T wave axis, and a decrease in QRS axis. For Parkinson's disease, the positive samples against the negative shows an increase of PR, T wave axis, as well as a decrease of P wave axis, QRS axis. For vascular dementia, positive samples against negative show an increase in the PR, QT, QTc, T wave axis, as well as a decrease in the QRS axis. For unspecified dementia, the positive samples against the negative shows an increase of PR, QT, QTc, as well as a decrease of RR, QRS axis. Finally, for delirium, the positive samples against the negative shows increase of QTc, T wave axis, as well as decrease of RR, PR, QT, QRS axis. Overall, the table highlights subtle but potentially clinically relevant ECG feature alterations associated with the investigated diagnoses, which may contribute to improved cardiac monitoring or risk stratification in these populations.

\end{document}